\documentclass[prc,preprint,superscriptaddress,showpacs,nofootinbib,tightenlines,amsmath]{revtex4}
\usepackage{graphics}
\usepackage{epsfig}
\usepackage{slashed}
\usepackage{pbox,calc}
\usepackage{braket}
\usepackage{bbm}
\usepackage{amsmath}
\usepackage{amssymb}

\usepackage{tikz}
\usetikzlibrary{intersections}
\usetikzlibrary{matrix}
\usetikzlibrary{trees}
\usetikzlibrary{calc}
\usetikzlibrary{decorations.pathmorphing}
\usetikzlibrary{decorations.pathreplacing}
\usetikzlibrary{decorations.markings}
\usetikzlibrary{3d}
\usetikzlibrary{external}

\pgfdeclaredecoration{complete sines}{initial}
{
  \state{initial}[
    width=+0pt,
    next state=sine,
    persistent precomputation={\pgfmathsetmacro\matchinglength{
        \pgfdecoratedinputsegmentlength / int(\pgfdecoratedinputsegmentlength/\pgfdecorationsegmentlength)}
        \setlength{\pgfdecorationsegmentlength}{\matchinglength pt}
    }] {}
  \state{sine}[width=\pgfdecorationsegmentlength]{
    \pgfpathsine{\pgfpoint{0.25\pgfdecorationsegmentlength}{0.5\pgfdecorationsegmentamplitude}}
    \pgfpathcosine{\pgfpoint{0.25\pgfdecorationsegmentlength}{-0.5\pgfdecorationsegmentamplitude}}
    \pgfpathsine{\pgfpoint{0.25\pgfdecorationsegmentlength}{-0.5\pgfdecorationsegmentamplitude}}
    \pgfpathcosine{\pgfpoint{0.25\pgfdecorationsegmentlength}{0.5\pgfdecorationsegmentamplitude}}
  }
  \state{final}{}
}

\tikzset{
  >=stealth,
  pion/.style=  {
                  dashed
                },
  pionarrow/.style=  {
                  dashed,
                  postaction={decorate},
                  decoration={markings, mark=at position .5 with {\arrow{>}}}
                },
  photon/.style=  {
                    decorate,
                    solid,
                    decoration={complete sines, amplitude=1mm, segment length=1.75mm, post length=0}
                  },
  rho/.style= {
                solid
              },
  rhoarrow/.style=  {
                  solid,
                  postaction={decorate},
                  decoration={markings, mark=at position .6 with {\arrow{>}}}
                }
}

\tikzset{->-/.style={decoration={
  markings,
  mark=at position #1 with {\arrow{>}}},postaction={decorate}}}
\tikzset{-<-/.style={decoration={
  markings,
  mark=at position #1 with {\arrow{<}}},postaction={decorate}}}

\begin{document}
\title{Axial-vector nucleon-to-delta transition form factors using the complex-mass renormalization scheme}
\author{Y. \"Unal}
\affiliation{Physics Department, \c{C}anakkale  Onsekiz Mart University, 17100 \c{C}anakkale, Turkey}
\author{A. K\"u\c{c}\"ukarslan}
\affiliation{Physics Department, \c{C}anakkale  Onsekiz Mart University, 17100 \c{C}anakkale, Turkey}
\author{S.~Scherer}
\affiliation{Institut f\"ur Kernphysik, Johannes Gutenberg-Universit\"at Mainz, D-55099 Mainz, Germany}
\date{April 7, 2021}

\begin{abstract}
    We investigate the axial-vector nucleon-to-delta transition form factors in 
the framework of relativistic baryon chiral perturbation theory at the one-loop order using the complex-mass renormalization scheme.
	We determine the available six free parameters by fitting to an empirical 
parametrization of the form factors obtained from the BNL neutrino bubble chamber experiments.  
	A unique feature of our calculation is the prediction of a non-vanishing 
form factor $C_3^A(Q^2)$.
	Moreover, our results show a surprising sensitivity to the coupling constant
$\texttt{g}_1$ of the leading-order Lagrangian ${\cal L}^{(1)}_{\pi \Delta}$. 
\end{abstract}

\maketitle

\section{Introduction}
   The $\Delta(1232)$ resonance is the first and best-established excitation of 
the nucleon \cite{Zyla:2020zbs}. 
   In the static quark model, it consists of three constituent quarks, coupled 
to spin $S=\frac{3}{2}$ and isospin $I=\frac{3}{2}$.
   The dominant decay mode by far is the strong decay into a pion and a 
nucleon, resulting in a lifetime of the order of 10$^{-23}$~s.
   According to Ref.~\cite{Zyla:2020zbs}, the pole position in the 
complex-energy plane is at $z_\Delta=m_\Delta-i\,\frac{\Gamma_\Delta}{2}\approx
(1210-i\,50)$~MeV.

    While there is a substantial amount of empirical information on the 
electromagnetic (vector) nucleon-to-delta transition \cite{Bartel:1968tw,Baetzner:1972bg,Stein:1975yy,Beck:1999ge,Pospischil:2000ad,
Mertz:1999hp,Joo:2001tw,Sparveris:2004jn,Elsner:2005cz,Kelly:2005jy,
Stave:2008aa,Aznauryan:2009mx,Blomberg:2015zma} (see, e.g., Refs.~\cite{Tiator:2011pw,Aznauryan:2011qj} for a review),
very little is known about the axial-vector nucleon-to-delta transition
\cite{Barish:1978pj,Radecky:1981fn,Kitagaki:1986ct,Kitagaki:1990vs,
Androic:2012doa}.
    The reason is twofold: (a) the weak probe couples only feebly to the 
nucleon-delta system and (b) the delta is unstable.\footnote{Of course, the second argument also applies to the electromagnetic transition. 
	Therefore, our knowledge of the electromagnetic transition form factors is
substantially less than for the nucleon elastic form factors.
    Concerning the transition from an unstable delta state to an unstable delta state,  Ref.~\cite{Zyla:2020zbs} quotes only a rough guess of the range, within which the magnetic moment is expected to lie.}
    On the theoretical side, numerous investigations exist for the 
electromagnetic case (see Ref.~\cite{Hilt:2017iup} and references therein) which have been extensively compared with data.
    Also, theoretical calculations of the axial-vector nucleon-to-delta 
transition have been performed in the framework of quark models \cite{Korner:1977rb,Hemmert:1994ky,Liu:1995bu,Golli:2002wy,
BarquillaCano:2007yk}, chiral effective field theory \cite{Zhu:2002kh,Geng:2008bm,Procura:2008ze,Unal:2018ruo}, lattice QCD \cite{Alexandrou:2006mc,Alexandrou:2009vqd,Alexandrou:2010uk}, and light-cone QCD sum rules \cite{Aliev:2007pi,Kucukarslan:2015urd}.

   While traditional calculations treat the delta resonance essentially as a
stable particle, it was emphasized in Ref.~\cite{Gegelia:2009py} that form factors of unstable particles should be determined from the renormalized 
three-point function at the complex pole.
   In fact, this idea was applied in Ref.~\cite{Hilt:2017iup} to the 
electromagnetic nucleon-to-$\Delta$ resonance transition to third chiral order in manifestly Lorentz-invariant chiral effective field theory.
   At the pole position, the magnetic dipole, electric dipole, and Coulomb
quadrupole form factors $G_M$, $G_E$, and $G_C$ are complex quantities.
   In particular, it was found that $G_E$ and $G_C$ have imaginary parts which 
are of the same magnitude as the respective real parts.
   In the present article, we extend the analysis to the axial-vector
transition at the one-loop level.  
   For that purpose, we combine a covariant description of the $\Delta(1232)$ 
resonance \cite{Hacker:2005fh,Wies:2006rv} with the complex-mass scheme (CMS)
applied to the chiral effective field theory of the strong interaction
\cite{Djukanovic:2009zn}.\footnote{
	The CMS was originally developed for deriving properties of $W$, $Z$, and
Higgs bosons obtained from resonant processes \cite{Stuart:1990,Denner:1999gp,Denner:2006ic,Actis:2006rc,Actis:2008uh}. }

   This article is organized as follows.
   In Sec.~II, we introduce the axial-vector nucleon-to-delta transition
process and discuss how it is related to weak pion production.
   In this context, we also define the pion-nucleon-delta form factor
   in terms of the PCAC relation (partially conserved axial-vector current).
   In Sec.~III, we present the effective Lagrangians we used.
   In Sec.~IV, we calculate the transition form factors and show our
results.
	Section V contains a comparison with other work.
	In Sec.~VI, we give a short summary.

\section{Axial-vector nucleon-to-delta transition form factors}
\subsection{Weak pion production}
\begin{figure}[htbp]
	\centering
	\includegraphics[width=0.5\textwidth]{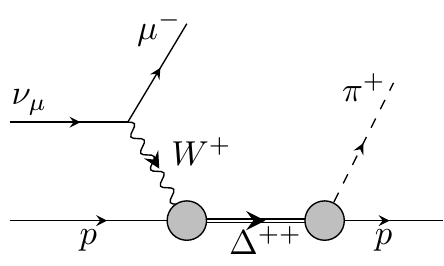}
	\caption{At $s\approx m_\Delta^2$,
		the process is dominated by the $s$-channel pole diagram due to the propagator of the $\Delta$(1232).
		The nucleon, electron, and neutrino are represented by single lines, the $\Delta$(1232) by a double line, the $W$ boson by a wiggly line, and the pion by a dashed line.
		The circles represent dressed vertices.}
	\label{fig:weak_pion_production}
\end{figure}

   The $\Delta$(1232) is an unstable particle with a very short lifetime of the
order of $10^{-23}$~s.
   Therefore, strictly speaking, stable one-particle states $|\Delta(p)\rangle$
with $p^2=m_\Delta^2$ do not exist \cite{Bjorken}.
   For this reason, direct measurements of transition form factors are
impossible, because the $\Delta$(1232) is not an asymptotic state of the strong interactions.\footnote{From a theoretical point of view, it is possible to study a hypothetical situation, where the sum of the nucleon and pion masses is larger than the $\Delta$ mass, resulting in a stable $\Delta$ state.}
   On the other hand, the existence of the delta is prominently seen in
pion-nucleon scattering or pion photoproduction on the nucleon.
   In other words, the impact of an unstable $\Delta$ may be investigated in
terms of a complete scattering amplitude, where it contributes
as an intermediate ``state.''
   In the present case, we are interested in the axial-vector nucleon-to-delta
transition.
   This may be studied in the weak production of a pion on the nucleon with 
hadronic center-of-mass energies in the delta region \cite{Adler:1968tw}. 
   For kinematical conditions such that the square root of the Mandelstam 
 variable $s$ is in the vicinity of the complex pole position,
 $$z_\Delta=m_\Delta-i\,\frac{\Gamma_\Delta}{2},$$
 the process is dominated by the propagation of a $\Delta$ resonance in the $s$ channel (see Fig.~\ref{fig:weak_pion_production}).  
    Since the $W$ boson induces the transition between the nucleon and the
delta in terms of the $V-A$ structure of the coupling to the quarks, this contribution is sensitive to both the nucleon-to-delta vector and axial-vector transitions.
	One now parametrizes the contribution of the unstable $\Delta$(1232) and 
defines the form factors in analogy to a stable particle.
	For an unstable particle such as the $\Delta$(1232), ``on-shell 
kinematics'' are given by the complex pole position.

\subsection{Definition of the axial-vector transition form factors}

   In the following, we provide a short definition of the form factors.
   We stick to the notation of Ref.~\cite{Unal:2018ruo}, where more details can
be found.
   In terms of the light-quark field operators, $q=(u,d)^T$, the Cartesian 
components of the isovector axial-vector current operator are given by \cite{Gasser:1983yg}
\begin{equation}
A^\mu_j(x)=\bar{q}(x)\gamma^\mu\gamma_5\frac{\tau_j}{2}q(x).
\end{equation}
   In general, the invariant amplitude\footnote{Our convention for the 
invariant amplitude complies with Ref.~\cite{Bjorken:1965zz}.
   In particular, it contains the imaginary unit on the right-hand side of 
Eq.~(\ref{invariant_amplitude}).} for a transition between hadronic states $|A(p_i)\rangle$ and $|B(p_f)\rangle$, induced by a plane-wave external field of the form $a_{\mu, j}(x)=\epsilon_{\mu, j}(q)e^{-iq\cdot x}$, is defined as
\begin{equation}
	\label{invariant_amplitude} 
	{\cal M}=i\epsilon_{\mu, j}(q)\langle B(p_f)|A^\mu_j(0)|A(p_i)\rangle,
\end{equation}
where four-momentum conservation $p_f=p_i+q$ due to translational invariance is implied.

   Introducing the spherical tensor notation \cite{Edmonds},
\begin{displaymath}
A^{\mu(1)}_{\pm 1}=\mp\frac{1}{\sqrt{2}}\left(A^\mu_1\pm i A^\mu_2\right),\quad A^{\mu(1)}_0=A^\mu_3,
\end{displaymath}
and using isospin symmetry, we express the matrix element of the spherical isospin components ($\alpha=+1,0,-1$) between a nucleon state and a $\Delta$ state as
\begin{equation}
\langle3/2,\tau_\Delta|A^{\mu(1)}_\alpha|1/2,\tau\rangle=
(1/2,\tau;1,\alpha|3/2,\tau_\Delta)\langle3/2||A^{\mu(1)}||1/2\rangle,
\end{equation}
where $\langle3/2||A^{\mu(1)}||1/2\rangle$ denotes the reduced matrix element and $(1/2,\tau;1,\alpha|3/2,\tau_\Delta)$ is the relevant Clebsch-Gordan coefficient.
   The reduced matrix element may, for example, be obtained from the $p$ to
$\Delta^+$ transition,
\begin{displaymath}
\langle 3/2||A^{\mu(1)}||1/2\rangle=\sqrt{\frac{3}{2}}\langle \Delta^+|A^{\mu(1)}_0|p\rangle.
\end{displaymath}

   The Lorentz structure of the reduced matrix element may be written as
\begin{equation}
	\label{Lorentz_structure}
	\langle \Delta(p_f,s_f)||A^{\mu(1)}(0)||N(p_i,s_i)\rangle
	=\bar{w}_\lambda(p_f,s_f)\Gamma^{\lambda\mu}_Au(p_i,s_i).
\end{equation}
   Here, the initial nucleon is described by the Dirac spinor $u(p_i,s_i)$ with 
a real mass $m_N$ and $p_i^2=m_N^2$, the final $\Delta$(1232) is described via the Rarita-Schwinger vector-spinor $\bar{w}_\lambda(p_f,s_f)$  \cite{Rarita:1941mf,Kusaka} with a complex mass $z_\Delta$ and $p_f^2=z_\Delta^2$ \cite{Gegelia:2009py,Agadjanov:2014kha}.
   In the following, it is always understood that the ``tensor''
$\Gamma^{\lambda\mu}_A$ is evaluated between on-shell spinors $u$ and $\bar{w}_\lambda$, satisfying\footnote{The explicit form of
$\bar{w}_\lambda$ can be found in Ref.~\cite{Agadjanov:2014kha}.}
\begin{align}
	\label{dirac}
	\slashed{p}_i u(p_i,s_i)&=m_N u(p_i,s_i),\\
	\label{rarita_schwinger}
	\bar{w}_\lambda(p_f,s_f)\slashed{p}_f&=z_\Delta \bar{w}_\lambda(p_f,s_f),\quad
	\bar{w}_\lambda(p_f,s_f)\gamma^\lambda=0,\quad
	\bar{w}_\lambda(p_f,s_f)p_f^\lambda=0.
\end{align}
   The expressions for a stable $\Delta$ resonance are obtained via the 
replacement $z_\Delta\to m_\Delta$.
   The ``tensor'' $\Gamma^{\lambda\mu}_A$ contains a superposition of four 
Lorentz tensors \cite{Adler:1968tw,LlewellynSmith:1971uhs}, which we choose as \cite{Alexandrou:2009vqd,Procura:2008ze}
\begin{align}
	\label{decomposition1}
	\Gamma^{\lambda\mu}_A&=\frac{C_3^A(Q^2)}{m_N}\left(g^{\lambda\mu}\slashed{q}-q^\lambda\gamma^\mu\right)
	+\frac{C_4^A(Q^2)}{m_N^2}\left(g^{\lambda\mu}p_f\cdot q-q^\lambda p_f^\mu\right)
	+C_5^A(Q^2)g^{\lambda\mu}+\frac{C_6^A(Q^2)}{m_N^2}q^\lambda q^\mu,
\end{align}
where $Q^2=-q^2$.
   Note that $p_f\cdot q=\frac{1}{2}(p_f+p_i+p_f-p_i)\cdot(p_f-p_i)
=\frac{1}{2}(p_f^2-p_i^2+q^2)=\frac{1}{2}(z_\Delta^2-m_N^2-Q^2)$.
   Our sign convention for the form factors in Eq.~(\ref{decomposition1}) is
such that we parameterize the matrix element of $+A^\mu_j$ and, thus, our sign convention follows closely the convention of the nucleon-to-nucleon axial-vector transition \cite{Schindler:2006it}.
   In particular, $C_5^A(Q^2)$ and $C_6^A(Q^2)$ correspond to the axial nucleon 
form factor $G_A(Q^2)$ and the induced pseudoscalar form factor $G_P(Q^2)$, respectively.

\subsection{Pion-nucleon-delta transition form factor}
   Assuming isospin symmetry, i.e., equal up-quark and down-quark masses 
$m_u=m_d=\hat m$, the divergence of the axial-vector current is given by  \cite{Scherer:2002tk,Schindler:2006it}
\begin{equation}
	\label{divergence_axial_current}
	\partial_\mu A^\mu_j(x)=\hat m P_j(x),
\end{equation}
where
\begin{equation}
	P_j(x)=i\bar{q}(x)\gamma_5\tau_j q(x)
\end{equation}
denotes the pseudoscalar density \cite{Gasser:1983yg}.
   With the help of the pion mass $M_\pi$ and the pion-decay constant $F_\pi$,
the isovector operator $\Phi_j(x)\equiv\hat m P_j(x)/(M_\pi^2 F_\pi)$ serves as an interpolating pion field \cite{Scherer:2002tk} such that Eq.~(\ref{divergence_axial_current}) amounts to the standard
PCAC relation (partially conserved axial-vector current) \cite{Adler-Dashen}.
   By means of $\Phi_j(x)$ we define the $\pi N\Delta$ transition form factor
$G_{\pi N\Delta}(Q^2)$ in analogy to the $\pi N$ form factor $G_{\pi N}(Q^2)$ \cite{Schindler:2006it,Gasser:1987rb} as \cite{Alexandrou:2009vqd}
\begin{equation}
\label{Pparametrization}
	\langle\Delta(p_f,s_f)||\Phi^{(1)}(0)||N(p_i,s_i)\rangle
	=i\frac{1}{M_\pi^2+Q^2} G_{\pi N\Delta}(Q^2)\bar w_\lambda(p_f,s_f)\frac{q^\lambda}{m_N}u(p_i,s_i).
\end{equation}
   From Eq.~(\ref{divergence_axial_current}) we obtain
\begin{equation}
	iq_\mu \langle \Delta(p_f,s_f)||A^{\mu(1)}(0)||N(p_i,s_i)\rangle=
	\hat m \langle \Delta(p_f,s_f)||P^{(1)}(0)||N(p_i,s_i)\rangle,
\end{equation}
which, using Eqs.~(\ref{decomposition1}) and (\ref{Pparametrization}), results in
\begin{equation}
	\label{GpiNDelta}
	G_{\pi N\Delta}(Q^2)
	=\frac{m_N}{F_\pi}\frac{M_\pi^2+Q^2}{M_\pi^2}
	\left[C_5^A(Q^2)-\frac{Q^2}{m_N^2}C_6^A(Q^2)\right].
\end{equation}
   In other words, once we know the form factors $C_5^A(Q^2)$ and $C_6^A(Q^2)$, 
we can also extract the form factor $G_{\pi N\Delta}(Q^2)$.
   The $\pi N\Delta$ coupling constant $g_{\pi N\Delta}$ is defined as
\begin{equation}
	\label{piNDeltacouplingconstant}
	g_{\pi N\Delta}=G_{\pi N\Delta}(-M_\pi^2).
\end{equation}
    Since the form factor $C_6^A(Q^2)$ has a pole at $Q^2=-M_\pi^2$, the 
coupling constant $g_{\pi N\Delta}$ does not vanish despite the factor $(M_\pi^2+Q^2)$ in Eq.~(\ref{GpiNDelta}).

\section{Effective Lagrangian}
\label{effective_lagrangian}
   In this section, we provide the interaction Lagrangians relevant for the 
calculation of the isovector axial-vector-current form factors of the nucleon-to-delta transition in covariant chiral EFT.
   The effective Lagrangian, $\mathcal{L}_\text{eff}$, consists of a purely 
pionic, a pion-nucleon, a pion-delta, and a pion-nucleon-delta Lagrangian,
each of which is organized in a combined derivative and quark-mass expansion.
   The most general effective Lagrangian for the calculation of the transition
form factors up to and including order $q^3$ is given by
\begin{equation}
	\mathcal{L}_\text{eff}=\mathcal{L}_{\pi}^{(2)}+\mathcal{L}_{\pi}^{(4)}
	+\mathcal{L}_{\pi N}^{(1)}+\mathcal{L}_{\pi \Delta}^{(1)}
	+\mathcal{L}_{\pi N \Delta}^{(1)}+\mathcal{L}_{\pi N \Delta}^{(2)}
	+\mathcal{L}_{\pi N \Delta}^{(3)}+ \ldots,
	\label{L}
\end{equation}
where the ellipsis denotes terms which are either of higher order or irrelevant for our calculation.

   The pionic Lagrangians at $\mathcal{O}(q^2)$ and $\mathcal{O}(q^4)$ are given by \cite{Gasser:1983yg,Scherer:2002tk}
\begin{equation}
\begin{aligned}
  \mathcal{L}_\pi^{(2)}=&\frac{F^2}{4}\left\{\text{Tr}[D_\mu U(D^\mu U)^\dagger]+\text{Tr}(\chi U^\dagger+U \chi^\dagger)\right\},\\
  \mathcal{L}_\pi^{(4)}=&\frac{l_4}{4} \text{Tr}[D_\mu U(D^\mu \chi)^\dagger]
  +D_\mu \chi (D^\mu U)^\dagger]+ \ldots,
\end{aligned}
\end{equation}
respectively, with $\chi=2B(s+ip)$, $s$ and $p$ denoting the scalar and pseudoscalar external sources \cite{Gasser:1983yg}.
   $F$ is the pion-decay constant in the chiral limit, 
$F_\pi=F[1+\mathcal{O}(\hat{m})]=92.2\;\text{MeV}$, and $B$ is associated with the scalar singlet quark condensate $\braket{\bar{q}q}_0$ in the chiral limit \cite{Gasser:1983yg,Scherer:2002tk,Colangelo:2001sp}.
   We employ SU(2) isospin symmetry, $m_u=m_d=\hat{m}$, and the lowest-order prediction for the pion mass squared is $M^2=2B\hat{m}$ \cite{Gasser:1983yg,Scherer:2002tk}, resulting
from inserting the quark masses into the external scalar field, $s=\hat{m}{\mathbbm 1}$.
   The triplet of pion fields is contained in the unimodular, unitary, 
$(2\times2)$ matrix $U$,
\begin{equation}
	U=u^2=\text{exp}\left(i\frac{\Phi}{F}\right), \quad
  	\Phi=\tau_j\phi_j=
	\begin{pmatrix}
  	\pi^0 &\sqrt{2}\pi^+  \\
  	\sqrt{2}\pi^- &-\pi^0
	\end{pmatrix},
\end{equation}
where $\tau_j$ are the Pauli matrices.
   Introducing external vector fields $v_\mu$ and axial-vector fields $a_\mu$ as
\begin{equation}
v_\mu=\frac{\tau_j}{2} v_{\mu, j},\quad
a_\mu=\frac{\tau_j}{2} a_{\mu, j},
\end{equation}
and using
\begin{equation}
  r_\mu=v_\mu+a_\mu,\quad
  l_\mu=v_\mu-a_\mu,
\end{equation}
  the covariant derivative of $U$ is defined as
\begin{equation}
   D_\mu U=\partial_\mu U-ir_\mu U+iU l_\mu.
\end{equation}

   The leading-order pion-nucleon Lagrangian reads 
\cite{Scherer:2002tk,Gasser:1987rb}
\begin{equation}
\label{LpiN1}
{\cal L}_{\pi N}^{(1)}
=\bar{\Psi}\left(i\slashed{D}-m+\frac{\texttt{g}_A}{2}\gamma^\mu\gamma_5 u_\mu\right)\Psi,\quad \Psi=\begin{pmatrix}p\\n\end{pmatrix},
\end{equation}
where $\Psi$ denotes the nucleon isospin doublet containing the four-component
Dirac fields for the proton and the neutron.
   The covariant derivative $D_\mu \Psi$ is given by\footnote{We do not 
consider a coupling to an external isoscalar vector field.}
\begin{equation}
\begin{split}
D_\mu\Psi&=(\partial_\mu+\Gamma_\mu)\Psi,\\
\Gamma_\mu&=\frac{1}{2}[u^\dagger(\partial_\mu-ir_\mu)u+u(\partial_\mu-il_\mu)u^\dagger].
\end{split}
\end{equation}
   The chiral vielbein is defined as
\begin{equation}
\label{vielbein}
u_\mu=u_{\mu,j}\tau_j=i[u^\dagger(\partial_{\mu}-ir_{\mu})u
-u(\partial_\mu-il_{\mu})u^\dagger].
\end{equation}
  The Lagrangian ${\cal L}_{\pi N}^{(1)}$ contains two free parameters, namely, 
the nucleon mass in the chiral limit, $m$, and the axial-vector coupling constant in the chiral limit, $\texttt{g}_A$.
   Expanding $u_{\mu, j}$ as
\begin{equation}
\label{vielbeinexpansion}
u_{j,\mu}=a_{\mu, j}-\frac{\partial_\mu\phi_j}{F}+{\cal O}(v_\mu\Phi,a_\mu\Phi^2,\partial_\mu\Phi\Phi^2),
\end{equation}
Eq.~(\ref{LpiN1}) gives rise to the lowest-order $\pi NN$ vertex 
as well as the axial-vector $NN$ transition vertex which are both needed for the one-loop corrections at ${\cal O}(q^3)$.

   The building blocks for constructing the Lagrangian of the $\Delta$ 
resonance can be found in Refs.~\cite{Hacker:2005fh,Scherer:2012zzd} and the references therein.
   For our purposes, we only need the leading-order contribution,\footnote{Note 
that the free Lagrangian contains an arbitrary real parameter $A\neq-\frac{1}{2}$ \cite{Moldauer:1956zz,Nath:1971wp}, for which we choose $A=-1$ such that the propagator takes the simplest form.}
\begin{equation}
\begin{split}
\mathcal{L}_{\pi \Delta}^{(1)}
&=-\bar{\Psi}_\mu\;\xi^\frac{3}{2}\Big[(i\slashed{D}-m_{\Delta})g^{\mu \nu}
-i(\gamma^{\mu}D^{\nu}+\gamma^{\nu}D^{\mu})+i\gamma^\mu
 \slashed{D}\gamma^\nu+m_{\Delta}\gamma^\mu \gamma^\nu\\
&\quad+ \frac{\texttt{g}_{1}}{2}(\slashed{u} \gamma_{5} g^{\mu\nu}
-\gamma^{\mu}u^{\nu}\gamma_{5}-u^{\mu}\gamma^{\nu}\gamma_{5}
-\gamma^{\mu}\slashed{u}\gamma_{5}\gamma^{\nu})\Big]\xi^\frac{3}{2}\Psi_\nu,
\label{LpiDelta1}
\end{split}
\end{equation}
from the $\pi\Delta$ Lagrangian,
where $\Psi_{\nu}$ denotes a vector-spinor isovector-isospinor field.
   The isovector-isospinor transforms under the
$1\otimes\frac{1}{2}=\frac{3}{2}\oplus\frac{1}{2}$ representation
and, thus, contains both isospin 3/2 and isospin 1/2 components.
   In order to describe the $\Delta$, it is necessary to project onto the
isospin-3/2 subspace.
   The corresponding matrix representation of the projection operator is denoted by
$\xi^\frac{3}{2}$, and the entries of $\xi^\frac{3}{2}$ are given by \cite{Scherer:2012zzd}
\begin{displaymath}
\xi^\frac{3}{2}_{ij}=\delta_{ij}-\frac{1}{3}\tau_i\tau_j.
\end{displaymath}   
   Inserting the expansion of Eq.~(\ref{vielbeinexpansion}) into Eq.~(\ref{LpiDelta1}), 
we obtain the leading-order $\pi\Delta\Delta$ vertex which is proportional to $\texttt{g}_{1}$ and which is needed for the one-loop corrections at ${\cal O}(q^3)$.

   The leading-order $\pi N\Delta$ chiral Lagrangian is given by [see Eq.~(4.200) of Ref.~\cite{Scherer:2012zzd} with $\tilde z=-1$ for consistency with the choice $A=-1$]
\begin{equation}
\label{LpiNdelta1}
{\cal L}^{(1)}_{\pi N\Delta}=\texttt{g}\bar{\Psi}_{\lambda,i}\xi^\frac{3}{2}_{ij}
(g^{\lambda\mu}-\gamma^\lambda\gamma^\mu)u_{\mu,j}\Psi+\text{H.c.},
\end{equation} 
where H.c.~denotes the Hermitian conjugate.
   Expanding $u_{\mu, j}$ as above,
Eq.~(\ref{LpiNdelta1}) gives rise to the leading-order contribution to $C_5^A(Q^2)$ as well as the leading-order $\pi N\Delta$ vertex, which is needed for the calculation of the loop contributions at ${\cal O}(q^3)$.

   At ${\cal O}(q^3)$, the higher-order Lagrangians 
${\cal L}^{(2)}_{\pi N\Delta}$ and ${\cal L}^{(3)}_{\pi N\Delta}$ can only contribute at the tree level.
   In principle, these Lagrangians were derived in Ref.~\cite{Jiang:2017yda}
   (see also Ref.~\cite{Holmberg:2018dtv}).
   Taking Eq.~(66) of Ref.~\cite{Jiang:2017yda} for 
${\cal L}^{(2)}_{\pi N\Delta}$, there would be no contribution to the form factors at ${\cal O}(q^2)$, because the first two terms contain the chiral vielbein quadratically, and the last term involves the "wrong" field strength tensor $f_+^{\mu\nu}$.
   However, as discussed in Appendix \ref{appendix_lagrangian}, there are
independent contributions at ${\cal O}(q^2)$.
   In fact, this is to be expected for the following reason. 
   Counting the polarization vector as of ${\cal O}(q)$ and treating only the
four-momentum $q$ (but not $p_f$) as a small quantity, we expect from Eqs.~(\ref{Lorentz_structure}) and (\ref{decomposition1}) two free parameters 
related to $C_3^A(0)$ and $C_4^A(0)$. 
   For the sake of simplicity, we will denote these two parameters by
   $\alpha$ and $\beta$, respectively (see Appendix \ref{appendix_lagrangian}).
   
   Similarly, the ${\cal O}(q^3)$ Lagrangian of Ref.~\cite{Jiang:2017yda}
produces fewer contributions to the form factors than is expected from the counting of momenta (and the polarization vector).
   Since it is not the purpose of this paper to construct the most general
Lagrangian at ${\cal O}(q^3)$, we have decided to Taylor expand the form factors and keep the expansion coefficients as free parameters.

\section{Results}
\begin{figure}[h]
\includegraphics[scale=0.85]{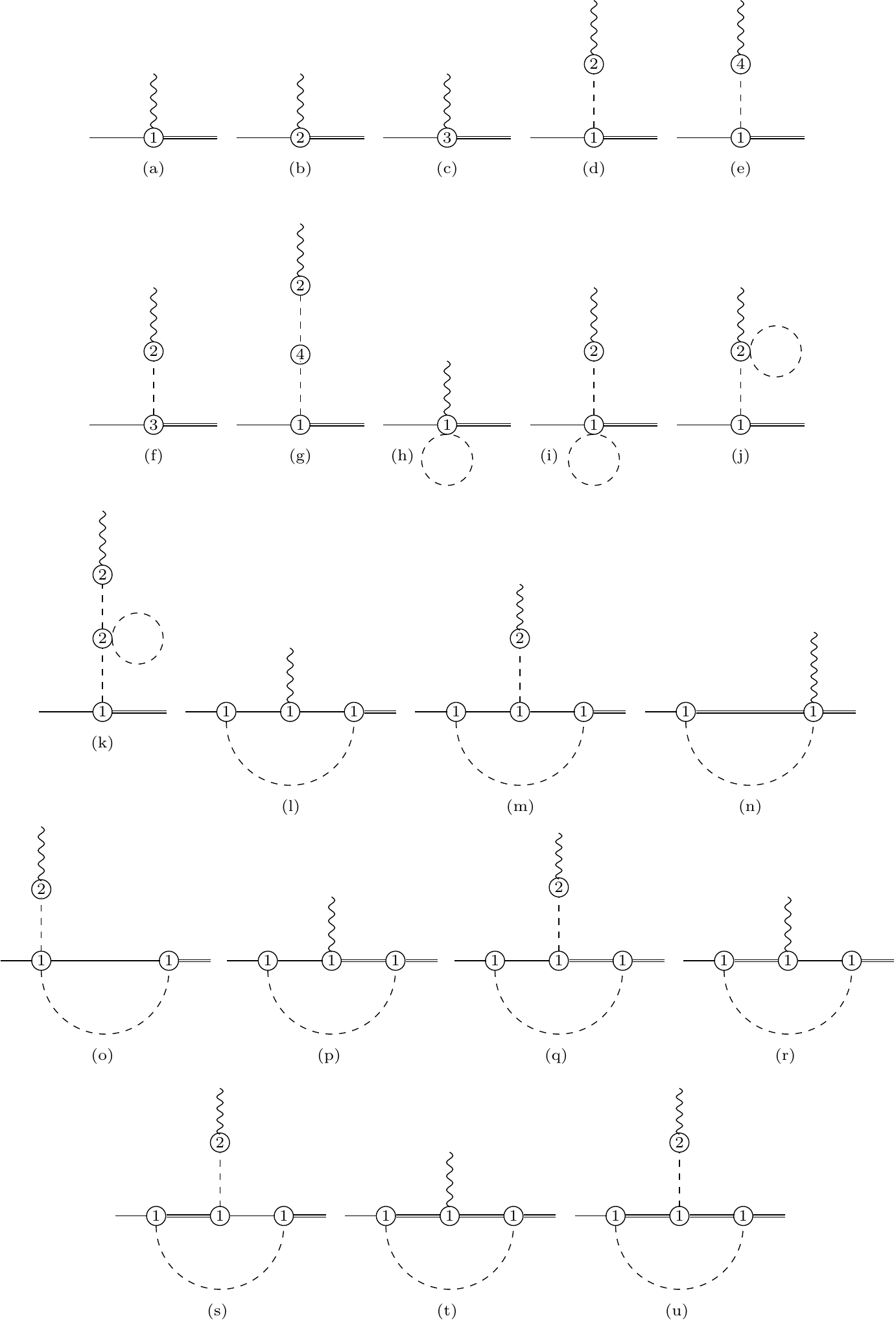}
\caption{Diagrams which, after renormalization, result in non-vanishing contributions to the axial-vector nucleon-to-delta transition form factors up to and including $\mathcal{O}(q^3)$.
The double, solid, dashed, and wiggly lines correspond to the delta, nucleon, pion, and the external axial-vector source, respectively.
\label{feynman_diagrams}}
\end{figure}
	Figure~\ref{feynman_diagrams} shows those tree-level and one-loop Feynman
diagrams that generate a non-vanishing contribution to the nucleon-to-delta transition matrix element of the isovector axial-vector current.
   In principle, the renormalized vertex is obtained by multiplying the
contributions of Fig.~\ref{feynman_diagrams} by the square roots of the wave function renormalization constants $Z_N$ and $Z_\Delta$.
   In practice, we evaluate the loop diagrams in the framework of dimensional
regularization at the renormalization scale $\mu=1$~GeV.
   We apply the modified minimal subtraction scheme of ChPT 
($\widetilde{\text{MS}}$) \cite{Scherer:2012zzd,Gasser:1983yg} by dropping infinite parts in terms of the combination $2/(n-4)-[\text{ln}(4\pi+\Gamma'(1)+1)]$, where $n$ denotes the number of space-time dimensions.
   We combine the remaining finite pieces with the available renormalized free
parameters. 

   In fact, for the actual calculation, we use a decomposition of 
$\Gamma^{\lambda\mu}_A$ which differs from Eq.~(\ref{decomposition1}),
namely,
\begin{equation}
\label{decomposition2}
\Gamma^{\lambda\mu}_A=k_1q^\lambda p_f^\mu+k_2 q^\lambda q^\mu+k_3q^\lambda \gamma^\mu+k_4 g^{\lambda\mu}.
\end{equation}
   For each diagram, we extract the four coefficients $k_i$ and determine their 
contributions to the form factors $C^A_i(Q^2)$, using the relations
\begin{align*}
C_3^A&=-m_N k_3,\\
C_4^A&=-m_N^2 k_1,\\
C_5^A&=\frac{1}{2}(z_\Delta^2-m_N^2-Q^2)k_1+(z_\Delta-m_N)k_3+k_4,\\
C_6^A&=m_N^2 k_2.
\end{align*}
	Using the strategy outlined in Sec.~\ref{effective_lagrangian}, the
tree-level contributions to the form factors can be written as\footnote{For the sake of simplicity, we omit the contribution of diagrams (e) and (g) in the formula for $C_{6\,\text{tree}}^A(Q^2)$.}
\begin{equation}
   	\label{tree_level_contributions}
   	\begin{split}
  	C_{3\,\text{tree}}^A(Q^2)&=\alpha,\\
  	C_{4\,\text{tree}}^A(Q^2)&=\beta,\\
  	C_{5\,\text{tree}}^A(Q^2)&=\texttt{g}+\gamma M^2+\delta Q^2,\\
 	C_{6\,\text{tree}}^A(Q^2)&=m_N^2\frac{\texttt{g}
   	+\gamma M^2+\epsilon M^2}{M^2+Q^2}+\zeta.
	\end{split}
\end{equation}
   We have explicitly shown the quark-mass dependence in terms of the
lowest-order squared pion mass $M^2$.
   In addition, we made use of the analogy of the form factors $C_5^A(Q^2)$ and
$C_6^A(Q^2)$ to the nucleon form factors $G_A(Q^2)$ and $G_P(Q^2)$ (see Ref.~\cite{Schindler:2006it}).
   At leading and next-to-leading order, the form factors $C_3^A(Q^2)$, 
$C_4^A(Q^2)$, and $C_5^A(Q^2)$ are constant.
   The parameters $\texttt{g}$, $\alpha$, and $\beta$ are independent
low-energy constants, i.e., they are not predicted by chiral symmetry.
   Using the relation from the static quark model with SU(6) symmetry
   results in the estimate $\texttt{g}=3\sqrt{2}\texttt{g}_A/5$
\cite{Unal:2018ruo,Hemmert:1997ye}.
   At this order, the form factor $C_6^A(Q^2)$ is predicted as    
\begin{displaymath}
   C_6^A(Q^2)=\frac{m^2_N}{M^2+Q^2}\texttt{g}.
\end{displaymath}  
   Turning to ${\cal O}(q^3)$, we will now have both tree-level modifications 
as well as loop contributions.
   The parameter $\gamma$ corresponds to a quark-mass correction to the 
nucleon-to-delta transition axial-vector coupling constant $g_{AN\Delta}\equiv G_5^A(0)$.
   Furthermore, $\delta$ contributes to the mean-square axial transition 
radius
\begin{equation}
\label{msar}
   \langle r^2_{AN\Delta}\rangle\equiv-\frac{6}{g_{AN\Delta}}
   \left.\frac{dG^A_5(Q^2)}{dQ^2}\right|_{Q^2=0}.
\end{equation}
	The parameter $\epsilon$ enters into the calculation of 
the generalized Goldberger-Treiman discrepancy \cite{Pagels:1969ne}
\begin{equation}
   \Delta\equiv 1-\frac{m_N g_{AN\Delta}}{F_\pi g_{\pi N\Delta}},
\end{equation}
where $g_{\pi N\Delta}$ is defined in Eq.~(\ref{piNDeltacouplingconstant}).  
   Finally, the parameter $\zeta$ is related to the $y$-intercept of $C_6^A(Q^2)$.   
 
   Unfortunately, there is very little empirical information on the form 
factors $C_i^A(Q^2)$.
   In order to obtain some estimate about the parameters and shape of our 
theoretical results, we perform a fit to the parametrization 
\begin{equation}
\label{ffparametrization}
C_i^A(Q^2)=\frac{C_i^A(0)\left(1+a_i\frac{Q^2}{b_i+Q^2}\right)}{\left(1
	+\frac{Q^2}{M_A^2}\right)^2}\quad(i=3,4,5)
\end{equation}
(see Appendix D of Ref.~\cite{Schreiner:1973mj}).
   The form factor $C_6^A(Q^2)$ is assumed to be dominated by the pion-pole 
contribution. 
   In particular, we make use of the axial-vector form factor parameters of the 
Adler model (see Table D.1 of Ref.~\cite{Schreiner:1973mj}), 
   \begin{equation}
    \begin{split}
    C_3^A(0)&=0,\quad a_3=0,\quad b_3=0,\\
    C_4^A(0)&=-0.3,\quad a_4=-1.21,\quad b_4=2.0\,\text{GeV}^2,\\
    C_5^A(0)&=1.2,\quad a_5=-1.21,\quad b_5=2.0\,\text{GeV}^2.
    \end{split}
    \end{equation}
    This parametrization was used by the authors of Ref.~\cite{Kitagaki:1990vs}
to extract the axial mass as $M_A=(1.28^{+0.08}_{-0.10})$~GeV from their analysis of the $\Delta^{++}$ production reaction $\nu_\mu+d\to
   \mu^-+\Delta^{++}+n_s$, where $n_s$ refers to a spectator neutron.

   To emphasize the low-$Q^2$ region, we choose the $Q^2$ values at
which to evaluate the empirical form factors according to $Q_n^2=n^2\times 0.004$ GeV$^2$ ($n=0,1,\ldots,28$).
   For the fits, we employ the {\it Mathematica} routine NonlinearModelFit.
   Our loop diagrams contain the low-energy constants (LECs) $\texttt{g}_A$, 
$\texttt{g}_1$, and $\texttt{g}$. 
   In the loop integrals, we approximate $\texttt{g}_A$ by $g_A=1.28$
(empirical value $g_A=1.2756\pm0.0013$ \cite{Zyla:2020zbs}),
because the difference between $\texttt{g}_A$ and $g_A$ is of order $M^2$ in the chiral expansion and, therefore, using $g_A$ introduces an error of higher order beyond the accuracy of our calculation.
   For the other two coupling constants, we make use of an SU(6) spin-flavor 
quark-model relation \cite{Scherer:2012zzd,Hemmert:1997ye},
\begin{equation}
\label{quarkmodelrelation}
\texttt{g}_1=\frac{9}{5}\texttt{g}_A,\quad
\texttt{g}=\frac{3}{5}\sqrt{2}\texttt{g}_A,
\end{equation}
resulting in $\texttt{g}_1=2.30$ and $\texttt{g}=1.08$, respectively.
   Furthermore, we replace $M^2$ by $M_\pi^2$ and use $M_\pi=0.135$~GeV.
   From a fit to the form factors $C_3^A(Q^2)$ and $C_5^A(Q^2)$, we obtain for 
the LECs $\beta$, $\gamma$, and $\delta$ the values
\begin{equation}
   \label{bgdfit1}
   \beta=0.335,\quad\gamma=-84.7\,\text{GeV}^{-2},\quad\delta=-3.20\,\text{GeV}^{-2}.
\end{equation}
   The value $C_5^A(0)$ is the analogue of $g_A$.
   The individual contributions to $C_5^A(0)$ are given by
\begin{equation}
   C_5^A(0)=1.20
   		   =\texttt{g}+\gamma M_\pi^2+\text{loops}=1.08 - 1.54 + 1.66.
\end{equation}
   The $M_\pi^2$ term plus the loops amount to a 10\% correction of the 
leading-order term.
   
\begin{figure}[h]
   	\includegraphics[width=\textwidth]{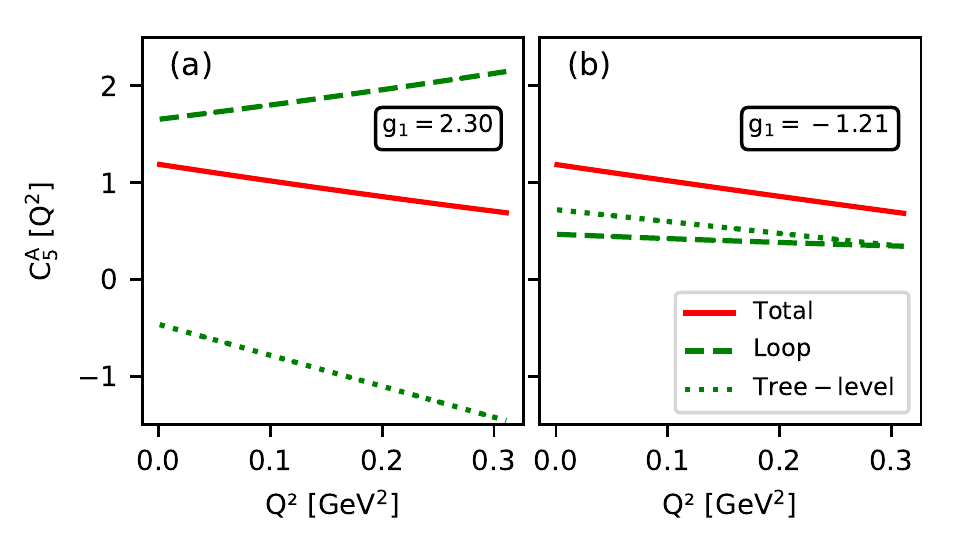}
   	\caption{\label{fig:C5A_individual} Individual contributions to $C_5^A(Q^2)$. The red solid, green dotted, and green dashed lines correspond to the total result, 
   		the loop contribution, and the tree-level contribution, respectively.
   	The left and right panels show the fits for $\texttt{g}_1=2.30$ and $\texttt{g}_1=-1.21$, respectively.}
   \end{figure}

\begin{figure}[h]
	\includegraphics[width=\textwidth]{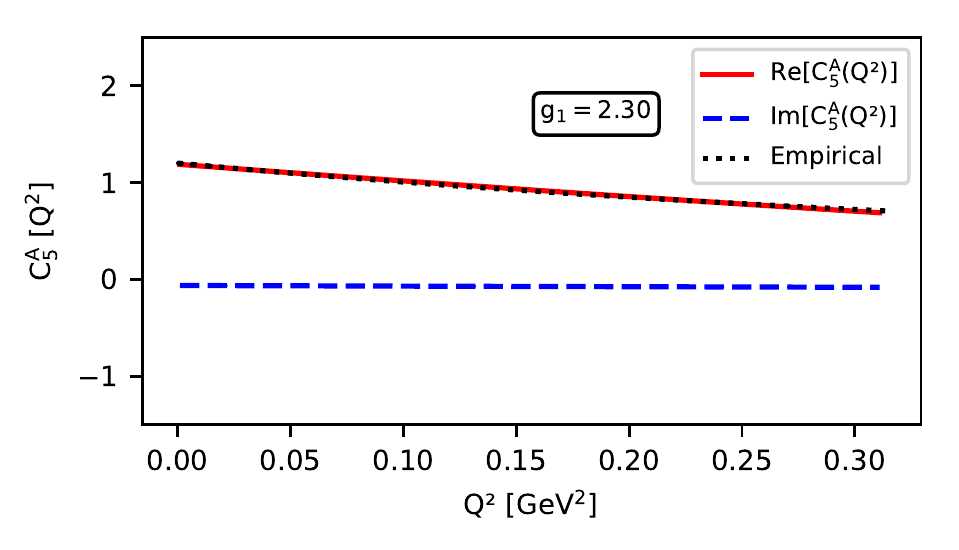}
	\caption{\label{fig:C5A_Fit} 
		Fit of the form factor $C_5^A(Q^2)$ for $\texttt{g}_1=2.30$. The red solid and blue dashed lines correspond to the real part and the imaginary part, respectively. The black dotted line is the empirical parametrization of the Adler model.}
\end{figure}

	In Fig.~\ref{fig:C5A_individual} (a), we display the individual 
contributions to $C_5^A(Q^2)$ for $\texttt{g}_1=2.30$; the total result is given by the red solid line, the loop contribution by the green dashed line, and the tree-level contribution by the green dotted line.
   The total result shows, as was to be expected from a calculation at 
${\cal O}(q^3)$, essentially a linear behavior as a function of $Q^2$.
   The loop contribution rises with increasing $Q^2$, whereas the tree-level 
contribution decreases linearly with $Q^2$.
	The two parameters $\gamma$ and $\delta$ are set in such a way that a good 
correspondence with the empirical parametrization of the Adler model was achieved (see Fig.~\ref{fig:C5A_Fit}).
	In the $Q^2$ range considered, the fit differs by less than 3\% from the 
empirical parametrization.

   Using Eq.~(\ref{msar}), we obtain for the mean-square axial 
transition radius
\begin{equation}
\label{resultmsatr}
\langle r^2_{AN\Delta}\rangle =0.345\, \text{fm}^2,
\end{equation}
which has to be compared with $\langle r^2_{AN\Delta}\rangle =0.427\, \text{fm}^2$ of the empirical parametrization.
   	A smaller axial radius for the fit was to be expected, because the
empirical parametrization contains more curvature while the fit behaves essentially linearly.
   Therefore, as can be seen from Fig.~\ref{fig:C5A_Fit}, the fit generates a slightly flatter 
behavior as a function of $Q^2$.

   As no reliable experimental data exist for $\langle 
r^2_{AN\Delta}\rangle$, we also quote for comparison the empirical values for the mean-square axial radius for the nucleon: 
$\langle r_A^2\rangle=(0.444\pm 0.018)$~fm$^2$ \cite{Bernard:2001rs}
and $\langle r_A^2\rangle=(0.453\pm0.023)$~fm$^2$ \cite{Bodek:2007ym}.
   In fact, based again on the quark-model relation of 
Eq.~(\ref{quarkmodelrelation}), the simplest assumption would be
$C^A_5(Q^2)=\frac{3}{5}\sqrt{2}G_A(Q^2)$ \cite{Unal:2018ruo}, resulting in
$\langle r^2_{AN\Delta}\rangle=\langle r_A^2\rangle$.

	Instead of using the quark-model estimate for $\texttt{g}_1$, we also made
use of the value $\texttt{g}_1=-1.21$ which was obtained in Ref.~\cite{Yao:2016vbz} from a fit to the $\pi N$ phase shifts of the $S$ and $P$ waves. 
	As stated in Ref.~\cite{Yao:2016vbz}, since $\texttt{g}_1$ appears only in 
the loop contribution of their calculation, a precise determination of its value is not to be expected.
	Note, in particular, that $\texttt{g}_1$ comes out with an opposite sign 
relative to the quark-model estimate.
	Using $\texttt{g}_1=-1.21$, we obtain for the LECs $\beta$, $\gamma$, and 
$\delta$ the values
\begin{equation}
	\label{bgdfit2}
	\beta=-0.0691,\quad\gamma=-19.8\,\text{GeV}^{-2},\quad\delta=-1.22\,\text{GeV}^{-2}.
\end{equation}
	Here, the individual contributions to $C_5^A(0)$ are given by
\begin{equation}
C_5^A(0)=1.19
=\texttt{g}+\gamma M_\pi^2+\text{loops}=1.08 - 0.36 + 0.47.
\end{equation}
    The individual contributions to $C_5^A(Q^2)$ for $\texttt{g}_1=-1.21$
are shown in Fig.~\ref{fig:C5A_individual} (b).
	In the range $0\leq Q^2\leq 0.3$~GeV$^2$, the total result for 
$\texttt{g}_1=-1.21$ deviates from that for $\texttt{g}_1=2.30$ by less than
1\%.
    However, the loop contribution behaves very differently, namely, at $Q^2=0$
it starts at a much lower value and it decreases with increasing $Q^2$ as opposed to a (stronger) increase with $Q^2$ for $g_1=2.30$.     
	In the end, this behavior is compensated by the rather different values of
$\gamma$ and $\delta$.
    
\begin{figure}[h]
	\includegraphics[width=\textwidth]{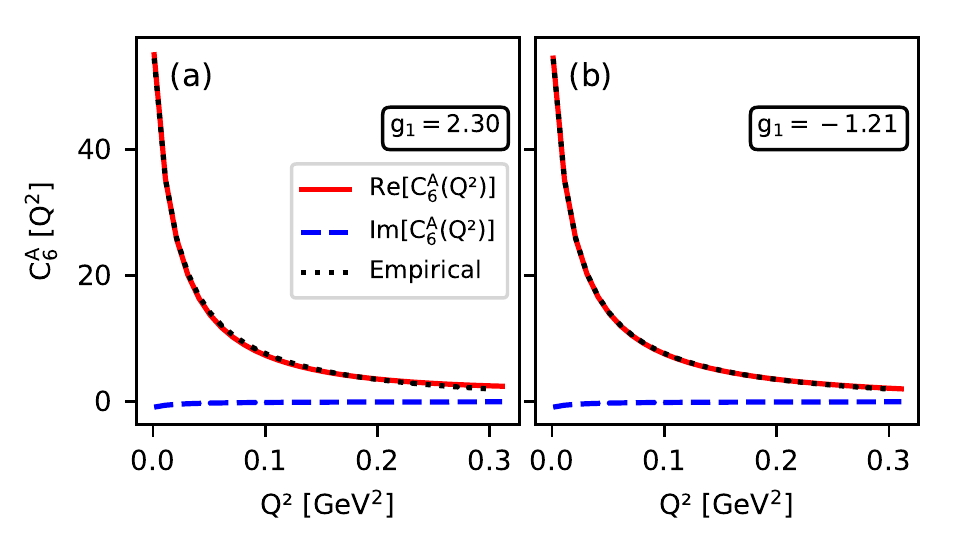}
	\caption{\label{fig:C6A_Fit} 
		Fit of the form factor $C_6^A(Q^2)$ for $\texttt{g}_1=2.30$ (left panel) and $\texttt{g}_1=-1.21$ (right panel), respectively. The red solid and blue dashed lines correspond to the real part and the imaginary part, respectively. The black dotted line is the 
		pion-pole-dominated 
		empirical parametrization with $C_5^A(Q^2)$ of the Adler model.}
\end{figure}

   The results for $C_6^A(Q^2)$ were obtained by fitting to the 
pion-pole-dominated expression,
\begin{displaymath}
   C_6^A(Q^2)=m_N^2 \frac{C_5^A(Q^2)}{M_\pi^2+Q^2},
\end{displaymath}
where $C_5^A(Q^2)$ is taken from Eq.~(\ref{ffparametrization}).
   Such a fit contains two free parameters, namely, $\epsilon$ and $\zeta$, 
which were obtained as
\begin{equation}
	\label{epsilon_zeta}
	\begin{split}
   \epsilon=59.3\, \text{GeV}^{-2}\quad\text{and}\quad\zeta=-11.5\quad&\text{for}\quad \texttt{g}_1=2.30,\\
   \epsilon=13.4\, \text{GeV}^{-2}\quad\text{and}\quad\zeta=-3.40\quad&\text{for}\quad \texttt{g}_1=-1.21.
   \end{split}
\end{equation}
   The corresponding results for $C_6^A(Q^2)$ are shown in
Fig.~\ref{fig:C6A_Fit}.
    For $Q^2\lessapprox 0.16$ GeV$^2$, the fit for $\texttt{g}_1=2.30$ is below
the empirical form factor and deviates by less than 2.6\% from the empirical 
$C_6^A(Q^2)$. 
    Beyond $Q^2\approx 0.16$ GeV$^2$, the fit is above the empirical result
with a continuously increasing deviation reaching $-23$\% at $Q^2=0.3$ GeV$^2$.
    For $\texttt{g}_1=-1.21$, the fit is generally closer to the empirical 
result than for $\texttt{g}_1=2.30$.
    For $Q^2\lessapprox 0.06$ GeV$^2$, the fit is below the empirical form 
factor, beyond $Q^2=0.06$ GeV$^2$ it is above.
    Again, the maximal deviation happens at $Q^2=0.3$ GeV$^2$ and amounts to
$-3.2$\%.    
   
\begin{figure}[h]
	\includegraphics[width=\textwidth]{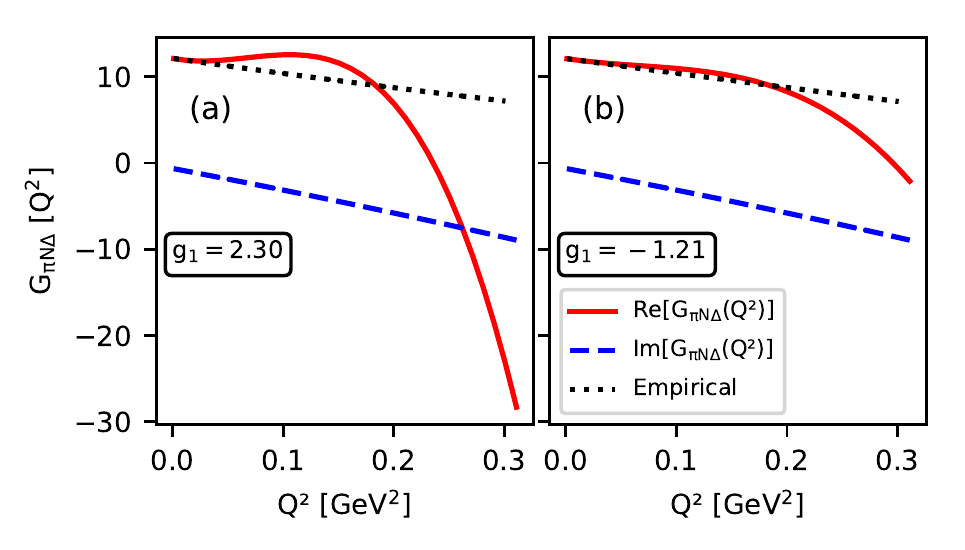}
	\caption{\label{fig:GpiND} 
		Results for the transition form factor $G_{\pi N\Delta}(Q^2)$ of Eq.~(\ref{GpiNDelta}) for $\texttt{g}_1=2.30$ (left panel) and $\texttt{g}_1=-1.21$ (right panel), respectively. The red solid and blue dashed lines correspond to the real part and the imaginary part, respectively. The black dotted line is the empirical result using
		$C_5^A(Q^2)$ and $C_6^A(Q^2)$ of the Adler model.}
\end{figure}

   Given the results for the form factors $C_5^A(Q^2)$ and $C_6^A(Q^2)$, we are 
now also in the position to discuss the $\pi N\Delta$ transition form factor
$G_{\pi N\Delta}(Q^2)$ of Eq.~(\ref{GpiNDelta}) (see Fig.~\ref{fig:GpiND}).
   For $\texttt{g}_1=2.30$, our result deviates very quickly with increasing
$Q^2$ from the pion-pole-dominated empirical result,
\begin{displaymath}
G_{\pi N\Delta}(Q^2)=\frac{m_N}{F_\pi}C_5^A(Q^2).
\end{displaymath}
   In fact, the linear combination of Eq.~(\ref{GpiNDelta}) involves a
delicate interplay between the terms 
\begin{displaymath}
\frac{m_N}{F_\pi}\frac{Q^2}{M_\pi^2}C_5^A(Q^2)\quad\text{and}\quad
-\frac{m_N}{F_\pi}\frac{Q^2}{m_N^2}\frac{M_\pi^2+Q^2}{M_\pi^2}C_6^A(Q^2).
\end{displaymath}
   The strong downward trend for increasing $Q^2\gtrapprox 0.15$ GeV$^2$ is due 
to a relatively large negative contribution proportional to $(Q^2)^3$, originating from the second term. 
 	The situation is somewhat better for $\texttt{g}_1=-1.21$, where the
deviation starts at $Q^2\approx 0.2$ GeV$^2$.
	This does not come as a surprise, because both $C_5^A(Q^2)$ and, in 
particular, $C_6^A(Q^2)$ are better described for $\texttt{g}_1=-1.21$.
	For the $\pi N\Delta$ coupling constant, we obtain
\begin{align*}
G_{\pi N\Delta}(-M_\pi^2)=g_{\pi N\Delta}&=12.8\quad\text{for}\quad
\texttt{g}_1=2.30,\\
g_{\pi N\Delta}&=12.5\quad\text{for}\quad
\texttt{g}_1=-1.21.
\end{align*}
    These values result in a generalized Goldberger-Treiman discrepancy of
\begin{equation}
\label{GGTD}
\Delta=0.0533\quad\text{and}\quad\Delta=0.0305,
\end{equation}
for $\texttt{g}_1=2.30$ and $\texttt{g}_1=-1.21$, respectively.
    Even though the imaginary parts of $C_5^A(Q^2)$ and $C_6^A(Q^2)$ are small,
we obtain a noticeable imaginary part for $G_{\pi N\Delta}(Q^2)$.
    In the present case, the imaginary part originates entirely from the loop
contributions.
   In this context, one should keep in mind that, in the complex-mass scheme, 
the low-energy constants can also be complex numbers.
   Therefore, in principle, they could generate additional imaginary 
contributions. 
   As our empirical ansatz for the form factors is real, we only fitted the 
real part of the form factors and left the imaginary tree-level contributions unspecified.
    
\begin{figure}[h]
	\includegraphics[width=\textwidth]{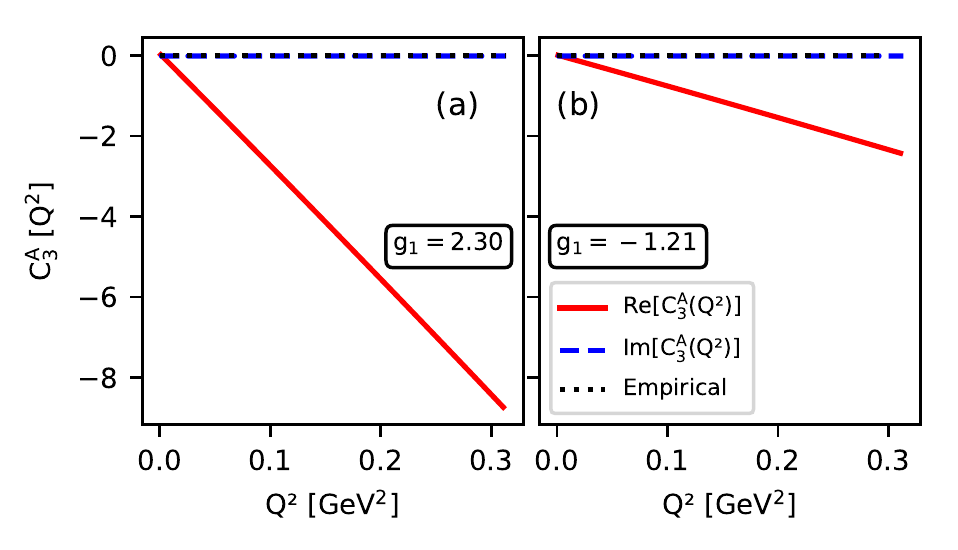}
	\caption{\label{fig:C3A_Fit} 
		Form factor $C_3^A(Q^2)$ for $\texttt{g}_1=2.30$ (left panel) and $\texttt{g}_1=-1.21$ (right panel), respectively. The red solid and blue dashed lines correspond to the real part and the imaginary part, respectively. In the Adler model, the form factor is set to zero (black dotted lines).}
\end{figure}

\begin{figure}[h]
	\includegraphics[width=\textwidth]{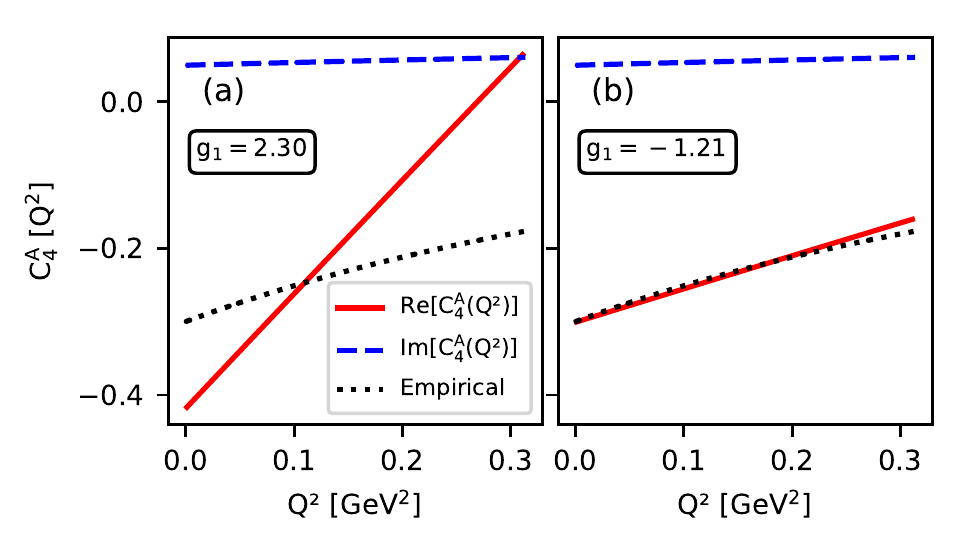}
	\caption{\label{fig:C4A_Fit} 
		Form factor $C_4^A(Q^2)$ for $\texttt{g}_1=2.30$ (left panel) and $\texttt{g}_1=-1.21$ (right panel), respectively. The red solid and blue dashed lines correspond to the real part and the imaginary part, respectively. The black dotted line is the empirical parametrization of the Adler model.}
\end{figure}

	Finally, we turn to the form factors $C_3^A(Q^2)$ and $C_4^A(Q^2)$. 
	These form factors have no analogue in the nucleon case.
	In Fig.~\ref{fig:C3A_Fit}, we show the loop contribution to $C_3^A(Q^2)$. 
	The parameter $\alpha$ of Eq.~(\ref{tree_level_contributions}) serves to 
shift the whole curve up or down and has been set to zero in the figure.
	No matter what the value for $\alpha$ is, our result is incompatible with 
the empirical ansatz $C_3^A(Q^2)=0$.
	By far the largest contribution to $C_3^A(Q^2)$ originates from diagram (t)
of Fig.~\ref{feynman_diagrams} and is proportional to $\texttt{g}\texttt{g}_1^2$.
   Therefore, the case $\texttt{g}_1=2.30$ produces a much stronger (negative) slope than 
$\texttt{g}_1=-1.21$.
	The result for $C_4^A(Q^2)$ is shown in Fig.~\ref{fig:C4A_Fit}.
	Here, we obtain a good description of the empirical form factor for 
$\texttt{g}_1=-1.21$, while the case $\texttt{g}_1$ again produces a much stronger (positive) slope.

\section{Comparison with other work}
	In this section, we provide a comparison of our calculation of the 
axial-vector current nucleon-to-delta transition form factors with other work.
	First, we compare our calculation with another calculation in the framework 
of covariant chiral perturbation theory \cite{Geng:2008bm}.
   	Our starting point is different in that we are using the 
isovector-isospinor representation for the $\Delta$ involving the projection operator $\xi$, whereas Ref.~\cite{Geng:2008bm} directly uses an isospin quadruplet.
	Furthermore, we have one more effective parameter than
Ref.~\cite{Geng:2008bm}, which affects the tree-level result of $C_6^A(Q^2)$.
    The parameters entering the loop diagrams are essentially the same, in 
particular, Ref.~\cite{Geng:2008bm} also uses the quark-model prediction
$H_A=\texttt{g}_1=\frac{9}{5}g_A$.
    Nevertheless, our loop contributions are, in general, substantially larger
in magnitude; this is particularly true for $C_3^A(Q^2)$.
	On the other hand, for both $\texttt{g}_1=2.30$ and $\texttt{g}_1=-1.21$, 
our total result for	$C_5^A(Q^2)$ is closer to the empirical parametrization than the result of Ref.~\cite{Geng:2008bm}, in particular for $Q^2\gtrapprox0.1$~GeV$^2$, whereas $C_6^A(Q^2)$ turns out to be very similar.
    For $C_3^A(Q^2)$ we obtain an opposite sign in comparison to 
Ref.~\cite{Geng:2008bm}.
    Finally, for $C_4^A(Q^2)$, an ambiguous situation arises. 
    For $\texttt{g}_1=-1.21$ there is a very good correspondence with the 
empirical form factor, whereas the result of Ref.~\cite{Geng:2008bm} is substantially below the empirical form factor.
    However, for $\texttt{g}_1=2.30$, our result increases too quickly, 
yielding too large a slope.
    In the framework of nonrelativistic chiral effective field theory to leading
one-loop order \cite{Procura:2008ze}, the $Q^2$ dependence of the form factors $C_5^A(Q^2)$ and $C_6^A(Q^2)$ to order three is entirely generated by counter terms and the pion-pole contribution, $C_3^A(Q^2)$ and $C_4^A(Q^2)$ gain $Q^2$ and $M_{\pi}$ dependence only at higher orders.
   Our values for the generalized Goldberger-Treiman discrepancy are slightly
larger than the $\sim 2\%$ predicted in the framework of heavy-baryon
chiral perturbation theory \cite{Zhu:2002kh}.
   The inclusion of the $a_1$ meson as an explicit dynamical degree of freedom 
was discussed in Ref.~\cite{Unal:2018ruo}.
   Besides the $a_1$-meson mass, this introduces one additional effective
parameter. 
   The $a_1$ meson effects the shape of the form factors $C_5^A(Q^2)$ and 
$C_6^A(Q^2)$; the form factor $C_5^A(Q^2)$ develops more curvature and 
$C_6^A(Q^2)$ lies above the pion-pole dominance prediction for $Q^2\gtrapprox
0.1$ GeV$^2$.
   Results from lattice QCD for $C_5^A(Q^2)$ and $C_6^A(Q^2)$ were reported in
Refs.\ \cite{Alexandrou:2009vqd,Alexandrou:2010uk}. 
   In general, the values of $C_5^A(0)$ in their calculations come out as 
smaller than one, i.e., they are also smaller than our results.
   At the same time, the axial mass turns out to be larger than $M_A=1.28$ GeV,
corresponding to a smaller mean-square axial transition radius. 
   This corresponds to our findings.
   At low $Q^2$, the lattice results for $C_6^A(Q^2)$ turn out to be smaller 
than our results.
   Finally, Ref.~\cite{Alexandrou:2010uk} contains two fits to 
$G_{\pi N\Delta}(Q^2)$.
   For the coupling constant $g_{\pi N\Delta}$, which
Ref.~\cite{Alexandrou:2010uk} defines at $Q^2=0$ rather than $Q^2=-M_\pi^2$, 
values between 8.44 and 16.3 are obtained, which have to be compared with our values 12.8 and 12.5.
   As a representative example for studies in the framework of nonrelativistic 
(chiral) quark models, we take the results of Ref.~\cite{BarquillaCano:2007yk}.
	First, we observe that the predictions of the form factors in the quark 
model are real quantities instead of complex functions in our calculation.
	The dominant form factor $C_5^A(Q^2)$ starts with $C_5^A(0)=0.93$ and
produces curvature according to an axial-vector meson dominance coupling (at the quark level) similar to Ref.~\cite{Unal:2018ruo}.
	The axial radius is predicted as $\langle r_{AN \Delta}^2\rangle=0.59$~fm$^2$ .
	They also predict a non-zero form factor $C_3^A(Q^2)$ in the low-$Q^2$ 
region, but of significantly smaller magnitude.
	Besides Ref.~\cite{Kitagaki:1990vs}, from which we took the empirical
parametrization of the form factors, some more recent experimental information is available for the linear combination
$G_{N\Delta}^A(Q^2)=\frac{1}{2}[m_N^2-m_\Delta^2+Q^2]C_4^A(Q^2)-m_N^2 C_5^A(Q^2)$,
which was extracted from the parity-violating asymmetry in inelastic electron-nucleon scattering near the $\Delta$ resonance.
	Reference \cite{Androic:2012doa} reports\footnote{Reference
\cite{Androic:2012doa} does not quote any units, even though in natural units the linear combination has dimension energy squared.} $G_{N \Delta}^A(Q^2 = 0.34~\text{GeV}^2)= - 0.05\pm (0.35)_{stat}\pm (0.34)_{sys}\pm (0.06)_{theory}$, whereas we obtained $-0.584$~GeV$^2$ for $\texttt{g}_1=2.30$ and $-0.544$~GeV$^2$ for $\texttt{g}_1=-1.21$.

\section{Summary}
   We analyzed the low-$Q^2$ behavior of the axial-vector nucleon-to-delta
transition form factors $C^A_3(Q^2)$, $C^A_4(Q^2)$, $C^A_5(Q^2)$, and $C^A_6(Q^2)$ at the one-loop level of relativistic baryon chiral perturbation theory.
   In total, the calculation involves six free parameters 
$\alpha$, $\beta$, $\gamma$, $\delta$, $\epsilon$, and $\zeta$ 
[see Eqs.~(\ref{tree_level_contributions})].
   The constants $g_A$, $F_\pi$, $M_\pi$, $m_N$, and $z_\Delta$ were fixed in 
terms of their empirical values.
   For the coupling constant $\texttt{g}$ we made use of the quark-model 
prediction $\texttt{g}=\frac{3}{5}\sqrt{2}g_A$.
   Finally, for the coupling constant $\texttt{g}_1$, appearing in certain loop
diagrams only, we considered two scenarios: we either made use of the quark-model prediction $\texttt{g}_1=\frac{9}{5}g_A=2.30$ or of $\texttt{g}_1=-1.21$ as obtained from an analysis of the $\pi N$ phase shifts of the $S$ and $P$ waves.
   Since there is essentially no direct experimental information on the form
factors available, we took the empirical parametrizations used in the analysis of Ref.~\cite{Kitagaki:1990vs} to determine our parameters.
	For our fits, we chose the interval $0\leq Q^2\leq 0.3136$ GeV$^2$,
where the upper end of the interval is likely to be at the verge of the applicability of a one-loop calculation.
    For the form factor $C_5^A(Q^2)$ we obtain good descriptions for both 
$\texttt{g}_1=2.30$ and $\texttt{g}_1=-1.21$, deviating from the empirical form by less than 3\% and 1\%, respectively.
	As can be seen from Fig.~\ref{fig:C5A_individual}, the loop corrections are 
sizable and their slope depends on the sign of $\texttt{g}_1$.
	As a consequence, the total result involves a delicate interplay between 
the loop contributions and the parameters $\gamma$ and $\delta$ [see Eqs.~(\ref{bgdfit1}) and (\ref{bgdfit2})].
	The parameters $\epsilon$ and $\zeta$ were determined from the fit to
$C_6^A(Q^2)$, where, again, for $\texttt{g}_1=-1.21$ our result is closer 
to the empirical form factor than for $\texttt{g}_1=2.30$ (see Fig.~\ref{fig:C6A_Fit}).
	As a result, the $\pi N\Delta$ transition form factor 
$G_{\pi N\Delta}(Q^2)$ deviates 
from the simple expectation $G_{\pi N\Delta}(Q^2)=\frac{m_N}{F_\pi}C_5^A(Q^2)$, again, more so for $\texttt{g}_1=2.30$ than for $\texttt{g}_1=-1.21$ 
(see Fig.~\ref{fig:GpiND}).
   For the $\pi N\Delta$ coupling constant we obtained $g_{\pi N\Delta}=12.8$ 
for $\texttt{g}_1=2.30$ and $g_{\pi N\Delta}=12.5$ for $\texttt{g}_1=-1.21$, resulting in the Goldberger-Treiman discrepancies $\Delta=0.0533$ and $\Delta=0.0305$, respectively. 
	The parameters $\alpha$ and $\beta$ are responsible for vertically shifting
the curves of the form factors $C_3^A(Q^2)$ and $C_4^A(Q^2)$, respectively, they cannot, however, modify their shapes.
	Therefore, the loop contributions are, to some extent, a unique 
feature of the predictions for $C_3^A(Q^2)$ and $C_4^A(Q^2)$.
   	In particular, our calculation predicts $C_3^A(Q^2)$ to be different from
zero in contrast to the empirical parametrization $C_3^A(Q^2)=0$
(see Fig.~\ref{fig:C3A_Fit}).
    Moreover, for $\texttt{g}_2=-1.21$ we obtain a very good agreement between 
our result for $C_3^A(Q^2)$ and the empirical form factor (see Fig.~\ref{fig:C4A_Fit}).
	A somewhat surprising feature is the fact that the negative value of
$\texttt{g}_1$ in all cases gives a better agreement with the empirical form factors than the quark-model result which uniquely predicts a positive sign.
	Unfortunately, as in the case of $\pi N$ scattering, $\texttt{g}_1$
does not enter the calculation at leading order but only at the loop level.	
	More about the sign of $\texttt{g}_1$ could possibly be learned from 
radiative pion-nucleon scattering $\pi N\to\pi\gamma N$ or radiative pion photoproduction $\gamma N\to \gamma \pi N$ in the $\Delta$-resonance region, where the $\pi\Delta\Delta$ vertex contributes at tree level and thus at leading order.

\section{Acknowledgments}
   The authors would like to thank J.~Gegelia for valuable discussions.
\begin{appendix}
	
\section{${\cal O}(q^2)$ Lagrangian}
\label{appendix_lagrangian}
   For our calculation, we need the pion-nucleon-delta interaction vertex and 
the axial-vector-nucleon-delta interaction vertex.
The building blocks that potentially contribute are the chiral vielbein
[see Eqs.~(\ref{vielbein}) and (\ref{vielbeinexpansion})] and
\begin{align}
\label{chiminus}
\chi_-&=u^\dagger\chi u^\dagger-u\chi^\dagger u\to M^2(U^\dagger-U)\to-2i M^2\frac{\tau_i\phi_i}{F},\\
\label{fminus}
f_{-\mu\nu}&=uf_{L\mu\nu}u^\dagger-u^\dagger f_{R\mu\nu}u\to f_{L\mu\nu}-f_{R\mu\nu}\to
-2(\partial_\mu a_\nu-\partial_\nu a_\mu)
=-\tau_i(\partial_\mu a_{\nu,i}
-\partial_\nu a_{\mu,i}).
\end{align}  
One also has to consider covariant derivatives of these building blocks. 

   According to Jiang {\it et al.} \cite{Jiang:2017yda}, the Lagrangian at 
${\cal O}(q^2)$, ${\cal L}_{\pi N\Delta}^{(2)}$, contains three structures [see Eq.~(66) of Ref.~\cite{Jiang:2017yda}].
The first two structures are proportional to the product $u_{\mu,i}u_{\nu,j}$ and thus contribute neither to the $\pi N\Delta$ interaction vertex nor to the
$aN\Delta$ interaction vertex.
The third structure is proportional to $f_{+\mu\nu}$ which contributes to the 
$vN\Delta$ interaction vertex but not to the $aN\Delta$ interaction vertex.
In other words, according to Jiang {\it et al.}, there are no contact 
interaction contributions to the transition form factors at ${\cal O}(q^2)$.

   Jiang {\it et al.}~compare their results with Ref.~\cite{Hemmert:1997ye},
which they quote as their Eq.~(67). 
   However, Hemmert {\it et al.}~\cite{Hemmert:1997ye} did not construct the
covariant version but rather the heavy-baryon version of the Lagrangian.
   According to Eq.~(82) of Ref.~\cite{Hemmert:1997ye}, they factor out 
$\exp(-iM_0 v\cdot x)$ with a common mass $M_0$ for both the nucleon field and the $\Delta$ field.
	The relevant heavy-baryon Lagrangian is then given in Eq.~(112), where $N$ 
and $T$ are heavy-baryon fields. 
	Note that there is no covariant Lagrangian in Ref.~\cite{Hemmert:1997ye}. In other words, Jiang {\it et al.}~must have reconstructed their Eq.~(67)	
from the heavy-baryon Lagrangian. 
	We make use of the results of section 5.5.~of
Ref.~\cite{Scherer:2002tk} to establish the connection. 
	Using Eq.~(5.122) of Ref.~\cite{Scherer:2002tk}, a single term $v^\mu$
originates from $\gamma^\mu$ and $2S^\mu$ from $\gamma^\mu\gamma_5$, respectively.

	Let us have a look at the first term of Eq.~(112) of Hemmert {\it et al.} This should result from
\begin{displaymath}
\frac{1}{2M_0}\bar{\Psi}^\mu_i b_1 i f_{+\mu\nu,i}\frac{1}{2}\gamma^\nu\gamma_5\Psi
+\text{H.c.}=\frac{1}{2M_0}\left(-\frac{1}{2}b_1i \bar{\Psi}^\mu_if_{+\mu\nu,i}\frac{1}{2}\gamma_5\gamma^\nu\Psi+\text{H.c.}\right),
\end{displaymath}
which, apart from a factor $1/(2M_0)$, agrees with the first term of Eq.~(67) of Jiang {\it et al.}\footnote{We left out the projector $O^{\mu\nu}_{A,n}(z_n)=g^{\mu\nu}
+(z_n+\frac{1}{2}(1+4z_n)A)\gamma^\mu\gamma^\nu$ as well as the projector
$\xi_{ij}^\frac{3}{2}$.}
   Now let us turn to the second term of Hemmert {\it et al.} which should
originate from
\begin{displaymath}
\frac{1}{2M_0}\bar{\Psi}^\mu_i ib_2 f_{-\mu\nu,i}\gamma^\nu\Psi
+\text{H.c.}
\end{displaymath}
   Comparing with Eq.~(67) of Jiang {\it et al.}, we see that they took the
wrong operator $D^\nu$ instead of $\gamma^\nu$ and then argue that such a term can be eliminated using arguments given in a previous section of 
Ref.~\cite{Jiang:2017yda}.
   In fact, Holmberg and Leupold \cite{Holmberg:2018dtv} also obtain a structure
analogous to the $b_2$ term in their construction for the decuplet-to-octet transition Lagrangian at next-to-leading order.
   
   We will now show that the $b_2$ term gives an explicit contribution to the 
form factor $C_3^A(Q^2)$ at ${\cal O}(q^2)$.
   Using Eq.~(\ref{fminus}) and after dropping the factor $1/(2M_0)$, from the $b_2$ term we obtain the Lagrangian 
\begin{displaymath}
ib_2\bar{\Psi}_{\lambda,i}\xi^\frac{3}{2}_{ij}(-\partial^\lambda a^\mu_j+\partial^\mu a^\lambda_j)\gamma_\mu\Psi+\text{H.c.},
\end{displaymath}
resulting in the form factor contribution
\begin{align*}
ib_2\bar{w}_\lambda\left((-(-iq^\lambda))\epsilon^\mu+(-iq^\mu)\epsilon^\lambda\right)\gamma_\mu u
=&ib_2\bar{w}_\lambda(iq^\lambda\gamma^\mu-i\slashed{q}g^{\lambda\mu})u\epsilon_\mu
=-b_2\bar{w}_\lambda(q^\lambda\gamma^\mu-\slashed{q}g^{\lambda\mu})u\epsilon_\mu\\
=&m_N b_2 \bar{w}_\lambda \frac{\slashed{q}g^{\lambda\mu}-q^\lambda\gamma^\mu}{m_N}u
\epsilon_\mu,
\end{align*}
from which we obtain the contribution $m_N b_2\equiv\alpha$ to the form factor $C_3^A(Q^2)$.

   Furthermore, we can relate the contribution to the form factor $C_4^A(Q^2)$ 
to the Lagrangian
\begin{equation}
\label{betaLagrangian}
\frac{\beta}{m_N^2}D_\mu \bar{\Psi}_{\lambda,i}\xi^\frac{3}{2}_{ij}f^{\lambda\mu}_{-j}\Psi+\text{H.c.},
\end{equation}
resulting in the invariant matrix element
\begin{displaymath}
{\cal M}=i\frac{\beta}{m_N^2}\epsilon_\mu\bar{w}_\lambda(g^{\lambda\mu} p_f\cdot q-q^\lambda p_f^\mu)u
\end{displaymath}
and, thus, the constant contribution $\beta$ to $C_4^A(Q^2)$.
   In fact, Holmberg and Leupold \cite{Holmberg:2018dtv} showed how to make use
of a total-derivative argument and the lowest-order equation of motion such that
the Lagrangian of Eq.~(\ref{betaLagrangian}) can be reexpressed in terms of the $b_2$ Lagrangian and terms of the ${\cal O}(q^3)$ Lagrangian.
    We will nevertheless stick to the Lagrangian of Eq.~(\ref{betaLagrangian}),
because, for our purposes, it is only relevant to know that we have a free parameter at our disposal, even if it originates from the ${\cal O}(q^3)$ and
which, only after rewriting, contributes to the transition matrix element in terms of $C_4^A(Q^2)$.
    Finally, it was shown in Ref.~\cite{Holmberg:2018dtv} that at 
${\cal O}(q^2)$ there is no ''new'' contribution to the $\pi N\Delta$ vertex.

\section{Loop integrals}
The scalar loop integrals of one-, two- and three- point functions which are used for the calculation of the loop diagrams are given by
\begin{align*}
&A_0(m^2)= \frac{(2\pi\mu)^{4-n}}{i\pi^2} \int  \frac{d^nk}{k^2-m^2+i0^+}, \\
&B_0(p^2, m_1^2, m_2^2)= \frac{(2\pi\mu)^{4-n}}{i\pi^2} \int \frac{d^nk}{[k^2-m_1^2+i0^+][(k+p)^2-m_2^2+i0^+]}, \\
&C_0(p_i^2, (p_f-p_i)^2, p_f^2, m_1^2, m_2^2, m_3^2) \\
&= \frac{(2\pi\mu)^{4-n}}{i\pi^2} \int \frac{d^nk}{[k^2-m_1^2+i0^+][(k-p_i)^2-m_2^2+i0^+][(k-p_f)^2-m_3^2+i0^+]}.
\end{align*}
\end{appendix}

\end{document}